\documentclass[trackchanges,twocolumn]{aastex701}

\newcommand{\hcop}{\mbox{HCO$^+$}}

\newcommand{\mum}{\mbox{$\mu$m}}
\newcommand{\kms}{\mbox{km\,s$^{-1}$}}

\newcommand{\sqcpg}{\mbox{cm$^{2}$\,g$^{-1}$}}
\newcommand{\cc}{\mbox{cm$^{-3}$}}
\newcommand{\lsol}{\mbox{$L_\odot$}}

\newcommand{\zsol}{\mbox{$Z_\odot$}}

\newcommand{\mujypbm}{\mbox{$\mu$Jy\,beam$^{-1}$}}

\newcommand{\hii}{\mbox{H\,{\sc ii}}}

\usepackage{comment}
\usepackage{CJK}

\begin{document}
\begin{CJK*}{UTF8}{gbsn}

\title{Magellanic Outflow and Chemistry Survey (MAGOS):\\
the Outflow-Luminosity Scaling Relations in the Large and Small Magellanic Clouds
}

\author[0000-0002-8389-6695]{Suinan Zhang (张遂楠)}
\affiliation{Department of Earth and Planetary Sciences, Institute of Science Tokyo, Meguro, Tokyo, 152-8551, Japan}
\affiliation{National Astronomical Observatory of Japan, 2-21-1 Osawa, Mitaka, Tokyo, 181-8588, Japan}
\email[show]{suinan.zhang@gmail.com}

\author[0000-0002-6907-0926]{Kei E. I. Tanaka}
\affiliation{Department of Earth and Planetary Sciences, Institute of Science Tokyo, Meguro, Tokyo, 152-8551, Japan}
\email[show]{kei.tanaka@eps.sci.isct.ac.jp}

\correspondingauthor{Suinan Zhang, Kei E. I. Tanaka}

\author[0000-0001-7511-0034]{Yichen Zhang}
\affiliation{Department of Astronomy, School of Physics and Astronomy, Shanghai Jiao Tong University, 800 Dongchuan Road, Shanghai 200240, People's Republic of China}
\affiliation{State Key Laboratory of Dark Matter Physics, School of Physics and Astronomy, Shanghai Jiao Tong University, Shanghai 200240, People's Republic of China}
\affiliation{Key Laboratory for Particle Astrophysics and Cosmology (MOE)/Shanghai Key Laboratory for Particle Physics and Cosmology, Shanghai 200240, People's Republic of China}
\email[]{}

\author[0000-0002-0095-3624]{Takashi Shimonishi}
\affiliation{Institute of Science and Technology, Niigata University, Ikarashi-ninocho 8050, Nishi-ku, Niigata 950-2181, Japan}
\email[]{}
%\email{shimonishi@env.sc.niigata-u.ac.jp}  

\author[0000-0002-8691-4588]{Yu Cheng}
\affil{National Astronomical Observatory of Japan, 2-21-1 Osawa, Mitaka, Tokyo, 181-8588, Japan}
\email[]{}

\author[0000-0002-2026-8157]{Kenji Furuya}
\affiliation{RIKEN Pioneering Research Institute, 2-1 Hirosawa, Wako-shi, Saitama 351-0198, Japan}
\email[]{}

\author[0000-0001-5817-6250]{Asako Sato}
\affiliation{Institut de Ciències de l'Espai (ICE-CSIC), Campus UAB, Can Magrans S/N, E-08193 Cerdanyola del Vallès, Catalonia, Spain}
\email[]{}

%\collaboration{all}{The MAGOS collaboration}

\begin{abstract}
Understanding massive star formation in low-metallicity environments is crucial for constraining the universality of star formation across cosmic evolution. 
In this Letter, we present the first systematic survey of massive protostellar outflows in the Large and Small Magellanic Clouds (LMC and SMC, $Z\sim$0.2--0.5~\zsol\/), as part of the Magellanic Outflow and Chemistry Survey (MAGOS). MAGOS targets 40 massive young stellar objects (MYSOs) with luminosities of (1--23)$\times 10^4$\lsol\/ with the Atacama Large Millimeter/submillimeter Array (ALMA) at 0.85~mm at a linear resolution of $\sim$0.1~pc. Using CO (3--2), we identify blue- and red-shifted outflows associated with the primary continuum sources toward 29 targets in the LMC and 10 targets in the SMC. Comparing the outflow mass, momentum, energy, and force versus luminosity scaling relations in MAGOS with those in the Galaxy,  
we find that the LMC CO outflows have a similar slope in the mass-luminosity relation, suggesting that CO remains an effective tracer of massive protostellar outflows at $Z\sim$0.5~\zsol\/. 
In contrast, despite the large scatter, the SMC CO outflows are deficient at high luminosity and show flatter scaling relations than the LMC and the Galaxy. This is consistent with reduced CO abundances due to enhanced photodissociation at high luminosity and low metallicity ($Z\sim$0.2~\zsol\/), although the possibility of intrinsically weak outflow launching cannot be ruled out with the current data.

\end{abstract}



\section{Introduction} \label{sec:intro}

% key science question
Massive stars have profound effects on the evolution of the interstellar medium (ISM) and galaxies across cosmic history. Despite recent advances in the understanding of massive star formation in the Galaxy \citep[see the review in, e.g.,][]{motte2018,beuther2025}, it remains poorly constrained how this process takes place in the early Universe.  
Metallicity is one of the key environmental factors distinguishing the present-day Universe from the early Universe. 
Exploring massive star formation in low-metallicity environments is thus central to understanding how it proceeds in the early Universe.

% theoretical work and observational discoveries and gap
Protostellar outflows, widely regarded as the signpost of ongoing mass accretion, are ubiquitously associated with low- and high-mass protostars in the Galaxy \citep[e.g.,][]{li2020, lu2021, hsieh2023, towner2024}. The masses, momenta, energies, and mechanical forces of outflows are correlated with bolometric luminosities over $\sim10^{-1}$--$10^6$~\lsol\/ \citep[e.g.,][]{beuther2002, wu2004, maud15b, towner2024}, indicating a consistent outflow driving mechanism across the mass spectrum. 
However, it is not yet clear whether the outflow-luminosity scaling relation established in the Galaxy holds in low-metallicity environments, raising questions about the universality of star formation processes across cosmic history. 

Theoretical studies predict that outflows driven by MHD disk winds can still be launched when the first adiabatic core forms even at $Z\gtrsim10^{-4}$~\zsol\/ \citep{higuchi2019}, and remain the dominant feedback mechanism regulating star formation efficiency at $Z\gtrsim$10$^{-1}$~\zsol\/ \citep{tanaka2018}. On the other hand, in low-metallicity environments where the dust opacity is reduced and cooling efficiencies are suppressed \citep[e.g.,][]{fukushima2020a}, molecules are more susceptible to photodissociation by UV radiation \citep[e.g.,][]{leroy2009}, which may affect the appearance of typical outflow tracers such as CO.  

Recent high-resolution infrared and submillimeter observations have discovered a handful of massive protostellar outflows in several nearby low-metallicity environments such as the outer Galaxy \citep[][]{cheng2025, ikeda2025} and the Large and Small Magellanic Clouds \citep[LMC/SMC;][]{fukui2015, shimonishi2016, tokuda2018,tokuda2022}. 
Despite the presence of massive outflows at $Z\sim$0.2--0.5\zsol\/ \citep[][]{russell1992, negueruela2015}, 
the rarity of detected outflows hinders quantitative studies on whether outflows in low-metallicity environments follow the same scaling relation with luminosity as those in the present-day Galaxy. 

% MAGOS survey
To systematically address this question, 
the Magellanic Outflow and Chemistry Survey (MAGOS; PI: K.\ Tanaka; Project code 2019.1.01770.S) was initiated with the Atacama Large Millimeter/submillimeter Array (ALMA). 
The LMC and SMC were selected because of their proximity \citep[$\sim$50--60~kpc;][]{pietrzynski2013, graczyk2020}, low metallicities \citep[$Z\sim$ 0.5~\zsol\/ for the LMC and $\sim$ 0.2~~\zsol\/ for the SMC,][]{russell1992}, and active star formation \citep{seale2009,carlson2012,oliveira2013, Fahrion2024}. 
40 spectroscopically confirmed massive young stellar objects (MYSOs) were observed with the ALMA at 0.85~mm at an angular resolution of $\sim$0\farcs3--0\farcs4 (equivalent to $\sim$0.1~pc). The sample includes 30 MYSOs in the LMC and 10 in the SMC, with a luminosity range of (1--23)$\times 10^4$\lsol\/. MAGOS has yielded fruitful discoveries of chemical properties of hot cores \citep{shimonishi2023, shimonishi2026} and filamentary clouds \citep{tokuda2023, tokuda2025} in low-metallicity environments. 

% introduce this letter
In this Letter, we present the outflow-luminosity scaling relations of the MAGOS sample, one of the key science goals of this survey. The Letter is organized as follows. The sample selection and observations are described in \autoref{sec:obs}. The employed 0.85~mm continuum and line data are summarized in \autoref{sec:results}. \autoref{sec:outflow_id} describes the identification of outflows and the calculations of outflow properties. \autoref{sec:relations} presents the outflow-luminosity scaling relations of the MAGOS sample and compares them with those of a Galactic sample. \autoref{sec:discussion} discusses the implications of our results for star formation in low-metallicity environments. Main conclusions are summarized in \autoref{sec:conclusions}.

\section{Sample and Observations} \label{sec:obs}

\begin{figure*}[t!]
\centering
\includegraphics[width=0.85\textwidth]{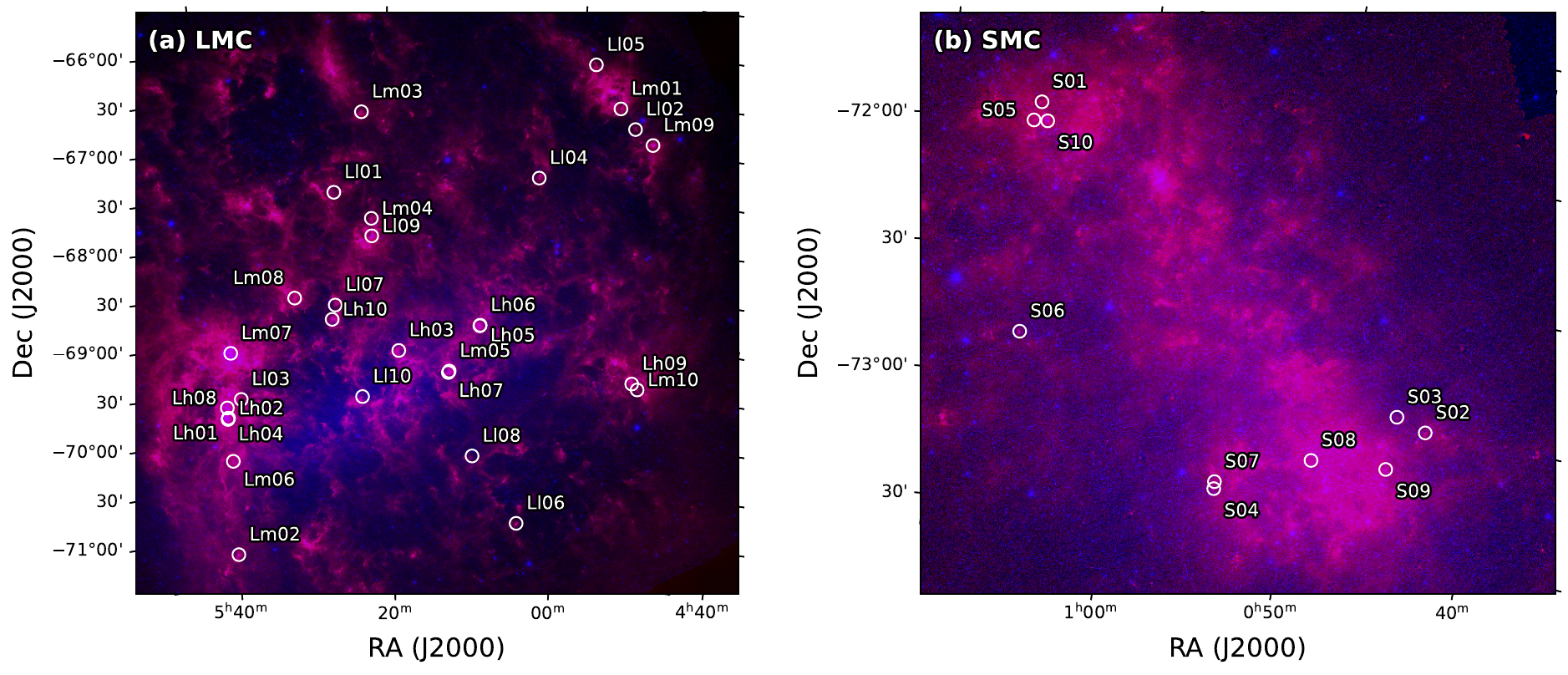}
\caption{Distribution of MAGOS targets. The background is a two-color composite image with the Herschel/PACS 160~\mum\/ emission \citep{meixner2010, meixner2013} in red and the Spitzer/IRAC 3.6~\mum\/ emission \citep{meixner2006, gordon2011} in blue. 
\label{fig:magos_gallery}}
\end{figure*}

The MAGOS sample is selected from the YSO catalogs in the Magellanic Clouds \citep{seale2009, oliveira2013} based on two criteria: (1) high luminosities ($\geq10^4$~\lsol\/) and (2) early evolutionary stages. 
%explain how the luminosities are estimated 
Bolometric luminosities of YSOs in the LMC are estimated with integrated fluxes over 1.2--500~\mum\/ by compiling the data from the Two Micron All Sky Survey \citep[2MASS,][]{skrutskie2006}, the Spitzer SAGE survey \citep{meixner2006}, and the Herschel HERITAGE survey \citep{meixner2010}. 
For SMC YSOs, the luminosities are estimated by fitting the spectral energy distribution (SED) over 0.3--24~\mum\/ \citep{oliveira2013, sewilo2013}. 
%explain evolutionary stage
YSOs are considered to be at early evolutionary stages based on their infrared spectral features such as silicate/ice absorption or polycyclic aromatic hydrocarbon (PAH) emission. 
With a total of 192 YSOs in the LMC satisfying the two criteria, we randomly select 10 YSOs for each of the three luminosity-divided groups: $10^4$--$10^{4.5}$~\lsol\/, $10^{4.5}$--$10^{5}$~\lsol\/, and $10^5$--$10^{5.5}$~\lsol\/, and refer to them as Ll01--Ll10, Lm01--Lm10, and Lh01--Lh10, respectively. 
For SMC, only 17 YSOs satisfy the selection criteria, so we randomly select 10 YSOs with luminosities of (1--14)$\times10^4$~\lsol\/, namely S01--S10. The YSOs are numbered in order of increasing luminosity. 
The spatial distribution of MAGOS targets is shown in \autoref{fig:magos_gallery}. 

%% The "ht!" tells LaTeX to put the figure "here" first, at the "top" next
%% and to override the normal way of calculating a float position.
%% The asterisk after "figure" tells the compiler to span multiple columns
%% if a two column style is selected.

We refer readers to \cite{shimonishi2026} for a detailed description of MAGOS observations and data reduction. Here we simply reiterate key parameters of the data employed in this work.  
The 0.85~mm continuum images have synthesized beam sizes ranging from 0\farcs35 $\times$ 0\farcs30 to 0\farcs38 $\times$ 0\farcs30 for the LMC and from 0\farcs36 $\times$ 0\farcs32 to 0\farcs37 $\times$ 0\farcs32 for the SMC. 
The position angles range from 11.80$^{\circ}$ to 38.58$^{\circ}$ for the LMC and from 19.48$^{\circ}$ to 32.80$^{\circ}$ for the SMC.   
The continuum images were constructed using bandwidths of 3.584--6.070~GHz and have rms noise levels before primary-beam correction of 1$\sigma$=100--140~\mujypbm\/ for the LMC, with bandwidths of 5.348--6.139~GHz and 1$\sigma$=60--70~\mujypbm\/ for the SMC.  
The CO (3--2) ($\nu_{\rm rest}=345.796$~GHz, $E_{\rm upper}=33.2$~K) and \hcop\/ (4--3) ($\nu_{\rm rest}=356.734$~GHz, $E_{\rm upper}=42.8$~K) cubes are restored to a common circular beam of 0\farcs4 ($\sim$0.1~pc) and have an rms noise of 1$\sigma$=0.3~K for the LMC and 1$\sigma$=0.2~K for the SMC. The channel width is 0.4~\kms\/. The maximum recoverable scale (MRS) is $\sim$4\farcs0 ($\sim$1~pc). Self-calibration was performed on several bright targets including Lh03--10, Lm07, Lm10, and Ll10.

\section{Results} \label{sec:results}
% Continuum source detection
% source nature: cross-match with HII region catalog
The 0.85~mm continuum images reveal bright continuum emission toward all targets except for Ll02, where no significant ($>5\sigma$) continuum emission is detected. 
We cross-match the MAGOS targets with compact \hii\/ region catalogs in the literature \citep{indebetouw2004,charmandaris2008,payne2004} and find that 8 targets are spatially associated with compact \hii\/ regions within 5\farcs\/ and 3 targets within 20\farcs\/. 
Toward these targets, the 0.85~mm continuum emission might be a combination of thermal dust emission and free-free emission \citep[e.g., Lh09,][]{nayak2019, cheng2026}. 

Notably, 13 targets have been resolved into $\geq$2 compact continuum sources separated by $\lesssim$1~pc (e.g., Lm04, Lh05),  
which are unresolved at far-infrared wavelengths (e.g., $\theta\sim$7\farcs\/ at Herschel/PACS 100\mum\/). 
These targets likely host multiple MYSOs instead of a single one. 
In this work, we focus on the brightest 0.85~mm continuum source for each target, because it is considered the primary contributor to the estimated bolometric luminosities, which are dominated by far-infrared fluxes. 

%line: SO/\hcop detection, CO detection.
Bright ($>$10~K) CO (3--2) emission is extensively detected toward all targets. High-velocity wing-like spectral features indicating outflowing gas are found toward 39 out of 40 targets.  
Ll02 is the exception, toward which the CO emission forms a ring-like structure with a velocity gradient. 
This source might be relatively more evolved, although it is not associated with any \hii\/ region in the literature to our knowledge, and is excluded from later analysis. 
In addition to CO (3--2), \hcop\/ (4--3) is another prominent molecular line detected toward all MAGOS targets. Its spatial distribution is well correlated with the continuum emission, suggesting that it effectively traces dense molecular gas \citep[$n(\rm H_2)\sim 10^5$~\cc\/,][]{tokuda2023}.  

\subsection{Outflow Identification}
\label{sec:outflow_id}

We employ CO (3--2) emission to identify protostellar outflows in the MAGOS sample. 
Although multiple outflows may be present when $\geq$2 compact continuum sources are resolved (e.g., Lm04, Lh05), only the outflows associated with the primary continuum source, which are also the strongest outflows in each field, are included in our analysis. 

The identification of outflows follows the strategy of the Galactic massive protostellar outflow survey in \cite{maud15b}. 
First, we search for high-velocity CO emission spatially associated with the primary continuum source through a channel-by-channel inspection of the CO cube. 
The maximum outflow velocity is determined when the CO emission spatially associated with the outflow drops below 5$\sigma_{\rm T_b}$, where $\sigma_{\rm T_b}$ is the rms noise of CO cubes. 
The minimum outflow velocity is determined by the full width at tenth maximum (FWTM, that is 1.8 $\times$ FWHM) of the average \hcop\/ spectrum of a $\sim0.6$~pc-diameter region centered on the continuum peak. 
This diameter is consistent with the aperture size for the targets at $\sim$6~kpc in \cite{maud15b}. 

\begin{figure*}[t!]
\centering
\includegraphics[width=0.95\textwidth]{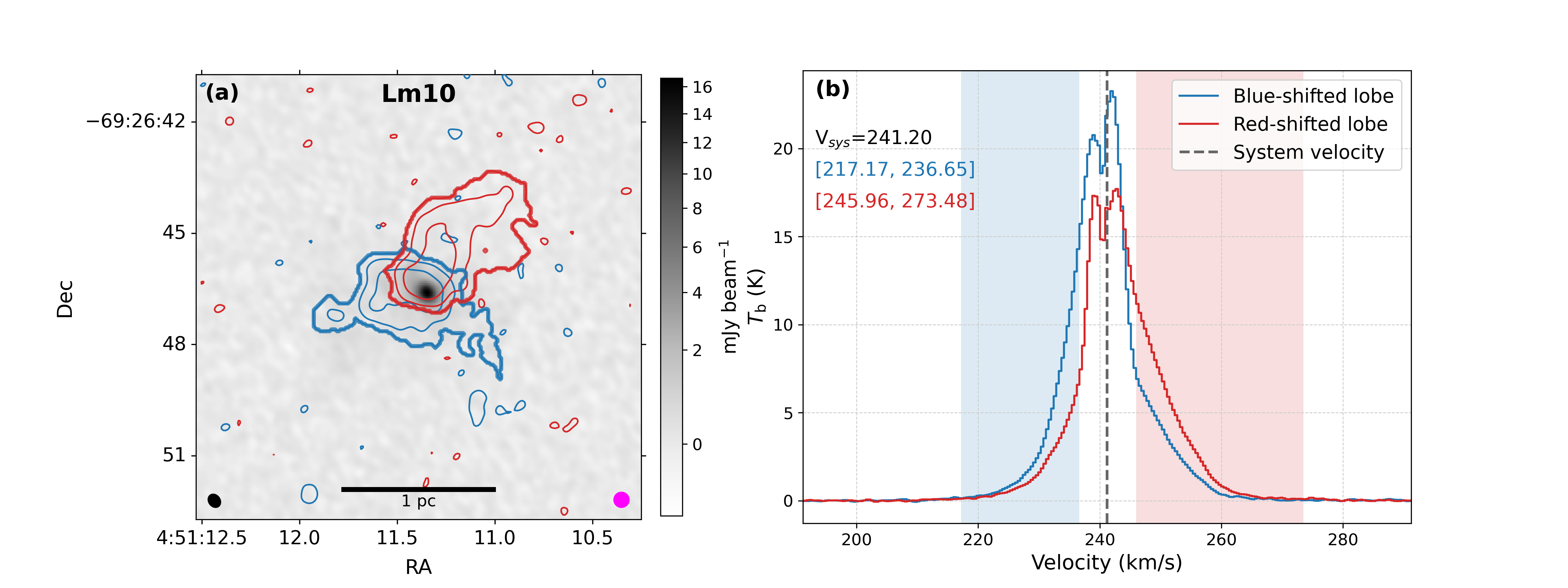}
\caption{
Example of outflows identified toward Lm10. (a) ALMA 0.85~mm continuum images (grayscale) overlaid with contours of the integrated blue-shifted (blue) and red-shifted (red) CO (3--2) emission. The contour levels are [1,15,25]$\times W_{\rm th}$ (see \autoref{sec:outflow_id} for details). 
Thick contours outline the determined outflow regions. 
The synthesized beams of the continuum image (black) and the CO cube (magenta) are shown in the lower left and right corners, respectively. 
A 1~pc scale bar is shown at the bottom. 
(b) CO (3--2) spectra averaged over the blue-shifted (blue) and red-shifted (red) outflow regions outlined by the thick contours in panel (a). The vertical dashed line indicates the systemic velocity $V_{\rm sys}=$241.20~\kms\/, while the shaded areas indicate the blue-shifted (blue) and red-shifted (red) outflow velocity intervals.  
\label{fig:lm10}}
\end{figure*}

Second, integrated blue- and red-shifted CO emission (moment 0) maps are generated with the outflow velocity extent. 
We adopt a threshold of the moment 0 map $W_{\rm th}$ to trace the spatial extent of outflows. The threshold $W_{\rm th}$ is defined as $W_{\rm th}=W_{\rm med}+5\times\sigma_{\rm MAP}$, where $W_{\rm med}$ is the median and $\sigma_{\rm MAP}$ is the rms noise that is given by $\sigma_{\rm MAP}=\sigma_{\rm T_b}\sqrt{N}\delta v$, where $N$ is the number of channels integrated over and $\delta v$ is the channel width in \kms\/. 
For $\sim60\%$ (23/39) of the targets, the $W_{\rm th}$ contours that are spatially associated with the primary continuum source well outline the spatial extent of primary outflows. For the remaining targets, their blue- and/or red-shifted integrated CO emission may be contaminated by (1) non-primary outflows and (2) extended CO emission not associated with outflows. Using $W_{\rm th}$ contours to trace the outflows of these targets frequently leads to overestimates of outflow properties for their primary continuum sources. 
Therefore, we instead manually select regions with high-velocity CO emission associated with the primary continuum source to highlight primary outflows for these targets. 
Targets for which the outflow regions are determined using contours or manual selected are noted in \autoref{tab:outflow_properties}. 
The identification of outflows is conducted with non-primary-beam-corrected cubes and moment 0 maps.  
\autoref{fig:lm10} shows an example of outflows identified toward Lm10, an MYSO with $8\times10^4$~\lsol\/ in the LMC. 
The outflows identified toward all MAGOS targets except for Ll02 are shown in \autoref{fig:outflow_gallery}. 

% show an example outflow Lm10

Outflow masses ($M$) are then derived from the primary-beam-corrected CO emission within the determined outflow areas using the equations below: 
\begin{equation}
M_{\rm lobe}=\sum_{(x,y)\in A_{\rm lobe}} \sum_{v\in V_{\rm lobe}}\delta M(x,y,v),
\end{equation}
where $\delta M(x,y,v)$ is given by
\begin{equation}
\delta M(x,y,v)=\delta A_{\rm pix}\, \alpha_{\rm CO}\, T_{\rm b}(x,y,v)\, \delta v,
\end{equation}
where $A_{\rm lobe}$ is the blue- or red-shifted outflow area, $\alpha_{\rm CO}$ is the CO(3--2)-to-H$_2$ conversion factor, $V_{\rm lobe}$ is the blue- or red-shifted outflow velocity range, $\delta A_{\rm pix}$ is the physical area of each pixel in pc$^{2}$, and $T_{\rm b}(x,y,v)$ is the CO brightness temperature in K at each voxel. 
Here we adopt a CO(3--2)-to-H$_2$ conversion factor of $\alpha_{\rm CO}$ = 10.4 M$_\odot$ pc$^{-2}$ (K km s$^{-1}$)$^{-1}$ for outflows in the LMC and $\alpha_{\rm CO}$ = 12.3 M$_\odot$ pc$^{-2}$ (K km s$^{-1}$)$^{-1}$ for the SMC. 
The adopted values are the typical $\alpha_{\rm CO}$ factors derived for the MAGOS sample using the CO (3--2) integrated emission and the gas mass inferred from the 0.85~mm continuum emission (see details in \autoref{sec:alpha}).  

Similarly, outflow momenta ($P$), energies ($E$), and forces ($F$) are derived by 
\begin{equation}
P_{\rm lobe}=\sum_{(x,y)\in A_{\rm lobe}} \sum_{v\in V_{\rm lobe}} \delta M(x,y,v)\,\left|v-V_{\rm sys}\right|, 
\end{equation}

\begin{equation}
E_{\rm lobe}=\frac{1}{2}\sum_{(x,y)\in A_{\rm lobe}} \sum_{v\in V_{\rm lobe}} \delta M(x,y,v)\,\left|v-V_{\rm sys}\right|^2,
\end{equation}

\begin{equation}
F_{\rm lobe}=\sum_{(x,y)\in A_{\rm lobe}} \sum_{v\in V_{\rm lobe}}\,\delta M(x,y,v)\,
\frac{\left|v-V_{\rm sys}\right|^2}{R(x,y)}, 
\end{equation}
where $V_{\rm sys}$ is the system velocity of the driving source and $R(x,y)$ is the distance of pixel $(x,y)$ from the location of the driving source, which is taken to be the location of its continuum peak. 
We apply a 5$\times\sigma_{\rm T_b}$ threshold to integrate only over voxels with high signal-to-noise ratios when deriving outflow properties. 
No inclination corrections are applied to the derived velocity-dependent outflow properties (see \autoref{sec:stats} for details).  
The derived outflow properties are summarized in \autoref{tab:outflow_properties}. 

\subsection{Outflow-Luminosity Relation} \label{sec:relations}

\begin{figure*}[t!]
\centering
\includegraphics[width=0.85\textwidth]{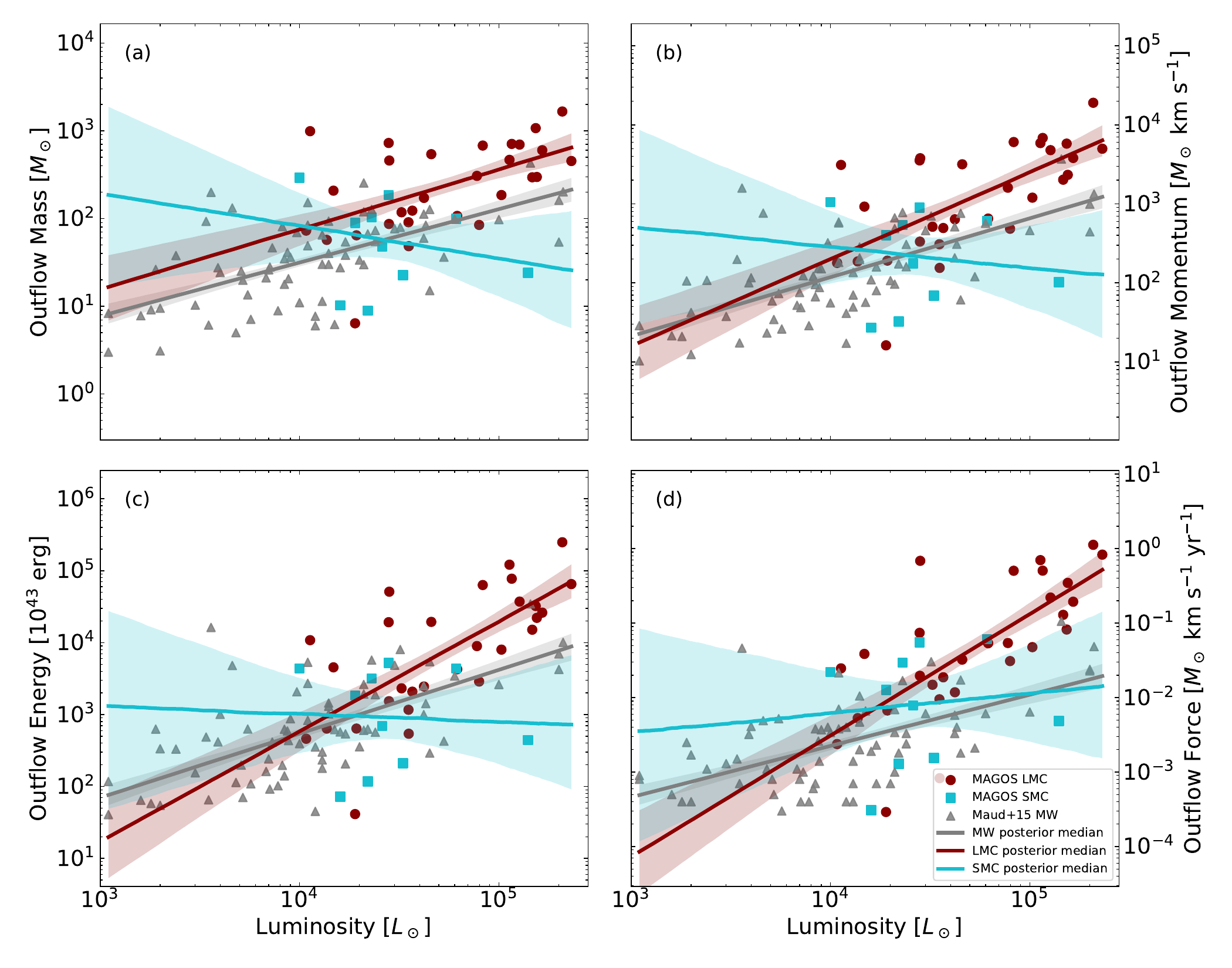}
\caption{Total outflow property versus luminosity scaling relations for MAGOS targets (LMC in red and SMC in blue) and the Galactic MYSOs (gray) at $\leq$6~kpc in \cite{maud15b}. The four panels show (a) outflow mass, (b) outflow momentum, (c) outflow energy, and (d) outflow force versus luminosity. Solid lines show the best-fitting Bayesian relations, with shaded areas indicating the 68\% credible interval.
\label{fig:outflow_luminosity_fit}}
\end{figure*}

% outflow propoerties vs luminosity relations
% statistics 
% linmix linear regression
We compare the relations of the outflow property versus luminosity for 39 MAGOS targets with those for the Galactic MYSOs with distances of $\leq$6~kpc in \cite{maud15b}. 
To quantify the outflow-luminosity scaling relations, we perform Bayesian linear regression to fit the LMC, SMC, and Galaxy data separately using a Python implementation of the linmix algorithm \citep{kelly2007, meyers2012}. 
The relation of the outflow property ($M$, $P$, $E$, $F$) versus luminosity ($L$) is described by:
\begin{equation}
    {\rm log_{10}~{\rm Property}} = \alpha+ \beta~({\rm log_{10}}L-{\rm log_{10}}L_0), 
\end{equation}
where $L_0$ is taken to be 10$^4$~\lsol\/. 
A total property of blue- and red-shifted outflows is adopted for each target. The fitting considers uncertainties in both outflow property and luminosity (see details in \autoref{sec:uncertainties}). 
\autoref{fig:outflow_luminosity_fit} shows the total outflow property versus luminosity for MAGOS and Galactic targets overlaid with posterior median scaling relations and 68\% credible intervals from Bayesian fits.

The LMC and Galactic outflows share a broadly similar slope in the mass-luminosity relation, with posterior median slopes of 
$\beta = 0.69^{+0.22}_{-0.21}$ and $\beta =0.61^{+0.10}_{-0.10}$, respectively. 
The uncertainty range represents the 68\% posterior credible interval. 
Notably, the LMC scaling relations are increasingly steeper than the Galactic relations in momentum, energy, and force, with differences between their posterior median slopes of $\Delta\beta_{\rm LMC-Gal}=$ 0.35, 0.65, and 0.95, respectively. 
In contrast, the SMC has shallower posterior median slopes than the LMC and the Galaxy in all properties, with the median slopes ranging from $\beta = -0.37^{+0.72}_{-0.73}$ in mass to $\beta = 0.28^{+0.98}_{-1.02}$ in force. 
In addition, the scaling relations of the SMC are less well constrained in all properties than those of the other two groups, as indicated by their considerably broader posterior distributions. 
Despite that, the 68\% credible band of the SMC slope remains below those of both the LMC and the Galaxy for mass and momentum. For energy, it remains below that of the LMC but slightly overlaps with the Galactic interval, while for force it remains below the LMC interval but substantially overlaps with the Galactic interval. 

% subsample tests
The variety in slope may reflect intrinsic differences in outflow scaling relations among the three groups, but could potentially be influenced by different sample sizes and luminosity ranges. 
To examine whether the deviations in the scaling relations in the LMC and SMC could arise purely from their small sample sizes and limited luminosity ranges, we perform 
Bayesian Galactic subsampling simulations matched to the LMC and SMC sample sizes, using Galactic sources with $L\geq10^4$~\lsol\/ (see details in \autoref{sec:stats}). 
The subsampling tests suggest that while a smaller sample size with a narrower luminosity range can yield a steeper slope with a larger scatter, it is insufficient to fully account for the observed deviations in the LMC and SMC outflow scaling relations. 

% discuss intercepts: whether the normalization differs?
% compare LMC-MW and SMC-MW separately.
% mention LMC-SMC separately
Similarly, we also compare the intercept $\alpha$, the normalization of the scaling relation at $L=10^4$~\lsol\/, among the three groups and the Galactic mock samples. 
As summarized in \autoref{tab:subsampling}, the posterior median intercepts for the MAGOS sample are higher than those for both the Galactic observed and mock samples. 
The excess is most pronounced for the LMC mass and momentum relations, while the broad SMC posterior intervals lead to substantial overlap with the corresponding mock distributions. 
In addition, the posterior median intercepts are higher in the SMC than in the LMC, but the broad and strongly overlapping 68\% posterior intervals indicate that the differences in normalization between the LMC and SMC are not well constrained. 

% conclusions
Overall, the linear regression analysis and subsampling tests suggest that the CO-traced outflow-luminosity scaling relations at $\geq10^4$~\lsol\/ in the LMC lie above those in the Galaxy in all properties and have a similar slope in mass and progressively steeper slopes in momentum, energy, and force. 
In contrast, outflow scaling relations in the SMC are flatter than those in the LMC and the Galaxy, mainly driven by deficient outflows in the high-luminosity regime. 

\section{Discussion \label{sec:discussion}}
% discuss observational bias and methods
\subsection{Deviations in Outflow Scaling Relations for the LMC and SMC \label{sec:deviation}}

The observed deviations in outflow scaling relations for the LMC and the SMC, despite essentially the same outflow identification strategy, might reflect environmental dependencies of CO-traced outflows, but before reaching robust conclusions one must first examine potential effects of the differences in underlying datasets and in the methods used to derive outflow properties. 

% resolution
The MAGOS CO data have a channel width of 0.4~\kms\/, identical to that of the Galactic survey. 
The linear resolution of the MAGOS data (0.1~pc) is also comparable to that of many of the nearer targets in the Galactic survey. 
% sensitivity
In addition, the MAGOS CO data ($1\sigma\sim$0.2--0.3~K) are more sensitive than the Galactic data ($1\sigma\sim$0.4--0.8~K, with a median of $\sim$0.7~K). 
Consequently, the exact threshold used to determine the maximum outflow velocity (5$\sigma_{\rm T_b}$) for the Galactic sample is on average $\sim$40\% higher than that adopted for the LMC and about a factor of two higher than that for the SMC. 
To test the sensitivity effect, we calculate variants of outflow properties for Lh09, the source driving the strongest outflows in the LMC, by increasing the threshold to match the typical threshold for the Galaxy.  
We find that applying a Galaxy-like threshold yields $<1\%$ difference in the derived outflow properties, indicating sensitivity differences may not play an important role. 

% spatial filtering - outflow size
With an MRS of $\sim$1~pc, the MAGOS data may miss smooth and extended emission due to spatial filtering, which should be fully recovered by the single dish observations of the Galactic sample. 
% casa simulation
To assess the effects of interferometric spatial filtering, we perform synthetic ALMA observations of the Galactic outflows associated with G111.5423$+$00.7776 \citep[NGC 7538 IRS1; $D\sim2.7$~kpc, $L\sim10^5$~\lsol\/;][]{maud15b}, using publicly available JCMT CO (3--2) data from the Perseus Arm Molecular Survey \citep[PAMS;][]{rigby2025}. The input CO cube is first scaled to the distances of the LMC and SMC, observed with MAGOS array configurations, and imaged using parameters matched to those of the MAGOS observations. By comparing the synthetic CO cubes with the input model, we find that approximately 77$\%$ and 80$\%$ of the total outflow mass is recovered for the synthetic LMC and SMC observations, respectively. 
Although limited to one Galactic example, these simulations suggest that spatial filtering may result in underestimates of outflow masses for the MAGOS sample. 
If a comparable fraction of CO emission is missing from the MAGOS observations, correcting for spatial filtering would increase the intercepts of the outflow scaling relations. 
The similar mass recovery fractions for the synthetic LMC and SMC observations also suggest that spatial filtering is unlikely to be the primary cause of the observed differences between their outflow-luminosity scaling relations. 

% higher intercepts of MAGOS: optical depth and alpha factor
On the other hand, when deriving outflow properties, \cite{maud15b} correct the optical depth of the $^{12}$CO emission per voxel using the optically thin $^{13}$CO emission, while we apply a uniform CO(3--2)-to-H$_2$ conversion factor across the outflow velocity range in this work (\autoref{sec:outflow_id}). 
The estimated $\alpha_{\rm CO}$ values can vary by a factor of a few from source to source (see \autoref{sec:alpha}).  
In particular, because $^{12}$CO emission can be optically thinner at higher velocities, a uniform $\alpha_{\rm CO}$ likely overestimates the outflow properties at higher velocities, leading to overestimates in mass, and progressively larger overestimates in momentum, energy, and force. 
This effect acts in the opposite direction to spatial filtering and may contribute to the $\lesssim$0.4~dex upward offset in the intercepts of the MAGOS scaling relations relative to the Galactic relations. 
However, the relative contributions of these two effects remain uncertain.

% steeper slopes in the LMC: steeper v_bulk-L relations due to a uniform alpha factor across velocity range
In addition, as discussed in \autoref{sec:relations}, the CO outflows in the LMC show progressively steeper scaling relations than those of the Galactic sample in momentum, energy, and force. 
This is consistent with a steeper outflow bulk velocity (given by $\langle v \rangle=\frac{P}{M}$) versus luminosity scaling relation in the LMC ($\beta=0.39\pm0.04$) than in the Galaxy ($\beta=0.12\pm0.04$). 
Because the LMC sources with higher luminosities generally have higher-velocity wings (see \autoref{tab:outflow_properties}), the adopted uniform $\alpha_{\rm CO}$ can exaggerate the importance of those wings, which steepens the $\langle v \rangle$-$L$ scaling relation in the LMC. 
Future observations targeting optically thin CO isotopes will help address this question. 

Overall, with our data it is most robustly established that (1) the LMC CO outflows have a similar scaling relation to that in the Galaxy in mass, and (2) the SMC CO outflows have flatter scaling relations than the LMC and the Galaxy.

\subsection{Implications for Star Formation in Low-metallicity Environments \label{sec:implication}}

Molecular clouds in lower-metallicity environments contain fewer dust grains. 
Reduced dust shielding and suppressed cooling efficiencies allow UV photons to penetrate deeper into clouds and propagate more freely \citep[e.g.,][]{fukushima2020a}. 
Consequently, molecules in low-metallicity clouds are more susceptible to photodissociation, resulting in lower molecular abundances \citep{leroy2009, shi2016}. 
The CO-traced protostellar outflows are thus expected to be fainter at a given luminosity in lower-metallicity environments. 

Our results, however, indicate a more complicated picture. 
% LMC
The similar mass-luminosity scaling relation between the LMC and the Galaxy suggests that substantial CO molecules survive photodissociation and remain as an effective tracer of protostellar outflows even around the most luminous MYSOs ($L\gtrsim10^{5}$~\lsol\/) at moderately low metallicity ($Z\sim$ 0.5~\zsol\/).  
%
% SMC
At a very low metallicity of $Z\sim$ 0.1--0.2~\zsol\/ in the SMC, CO outflows start to become deficient, particularly in the high luminosity regime. 
The three brightest targets in the SMC (S08--10; $L\gtrsim10^{4.5}$~\lsol\/) are spatially associated ($<$10\farcs\/) with compact \hii\/ region candidates \citep{payne2004, cotton2024}, implying the presence of copious UV photons emitted by either the central MYSOs driving outflows or adjacent MYSOs. 
The strong UV radiation in a very low-metallicity environment can further strengthen CO photodissociation, reducing its abundance and detectability.  
The observed deficient CO outflows at high luminosity in the SMC are thus consistent with enhanced photodissociation as a combined result of strong UV radiation and a very low metallicity. 

Alternatively, the deficient CO outflows in the SMC might reflect intrinsically weak outflow launching. 
Theoretical studies predict that outflows driven by magnetocentrifugal disk winds, one of the widely accepted outflow launching mechanisms \citep{pudritz2019}, operate even at extremely low metallicity \citep[e.g., $Z\gtrsim10^{-4}$~\zsol\/,][]{tanaka2018, higuchi2019}. 
Nonetheless, no clear predictions have been made on the metallicity dependence of outflow properties. 
Instead, outflow strengths are suggested to strongly depend on magnetic fields and accretion rates \citep{matsushita2017, machida2021}.  
\cite{machida2021} demonstrate that the combination of weaker magnetic fields and higher accretion rates produces weaker outflows, but it is unclear whether this still holds in low-metallicity environments. 

\begin{figure}[t!]
\centering
\includegraphics[width=0.4\textwidth]{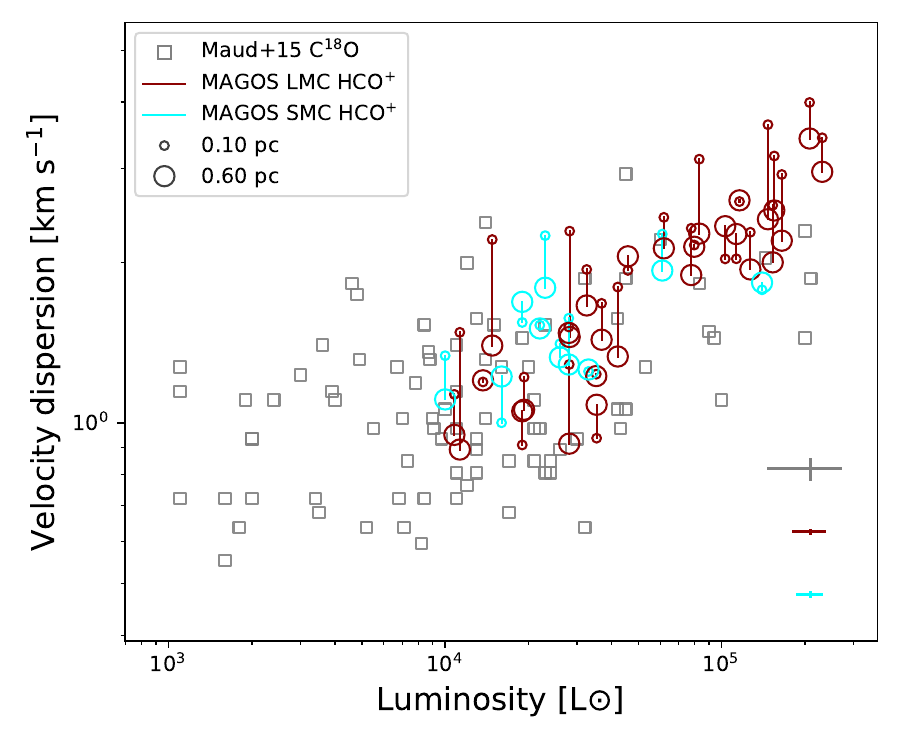}
\caption{Velocity dispersion versus luminosity for the MAGOS LMC (red) and SMC (blue) MYSOs and the Galactic MYSOs (gray) at $\leq$6~kpc in \cite{maud15b}. Vertical lines connect velocity dispersions estimated over 0.1~pc and 0.6~pc apertures for each MAGOS target as indicated by marker sizes. The aperture sizes are selected to be comparable with those for the Galactic sources. Typical uncertainties for each group are shown in the lower right corner (see \autoref{sec:uncertainties} and \citealt{maud15a}). 
\label{fig:dv_luminosity}}
\end{figure}

% sigma-L relation: LMC/SMC feedback at 0.1--0.6 pc scale
% 0.6 pc is the aperture size of the most distant Maud+15 source
As an additional constraint on the gas kinematics in the immediate surroundings ($<1$~pc) of MYSOs in different environments, we examine the velocity dispersion versus luminosity relation for the MAGOS and the Galactic samples using dense gas tracers. 
As shown in \autoref{fig:dv_luminosity}, the velocity dispersion versus luminosity relation for MAGOS targets is broadly consistent with that of Galactic sources, indicating that the dense gas kinematics around sources driving CO outflows are not drastically different across environments. 
This does not, however, rule out the possibility of intrinsically weak outflow launching, because the observed velocity dispersions around MYSOs at $\sim$0.1--0.6~pc scales are likely the combined result of, e.g., gravitational infall, rotation, turbulent motions, and protostellar feedback \citep[e.g.,][]{peretto2023}. 
This is also supported by the Galactic outflow study, where no clear evidence is found that outflows dominate the kinetic energy of dense cores at scales of $\sim$0.4--2~pc \citep{maud15a,maud15b}. 

Overall, our results are consistent with a joint dependence of CO outflows on metallicity and luminosity:  
at a moderately low metallicity of $Z\sim$ 0.5~\zsol\/, substantial CO molecules remain undestroyed even at high luminosity ($L\gtrsim10^{5}$~\lsol\/); at a very low metallicity of $Z\sim$ 0.1--0.2~\zsol\/, photodissociation becomes increasingly effective at higher luminosity (e.g., $L\gtrsim10^{4.5}$~\lsol\/), making outflows underluminous in CO. 
In this scenario, while the chemical behaviors of star formation processes may be regulated by the environment (i.e., metallicity and radiation field), the physical processes in star formation (e.g., outflow launching) might remain universal across cosmic history. 
However, with the current data the possibility of deficient CO outflows directly due to intrinsically weak outflow launching cannot be excluded. 
If this is the case, it would challenge the universality of basic physical processes in star formation across cosmic history. 
% future work
Future observations targeting outflow tracers of CO-dark gas are essential to distinguish between CO photodissociation and intrinsically weak outflow launching. 

\section{Conclusions} \label{sec:conclusions}

We present the first systematic survey of massive protostellar outflows for 40 MYSOs with $L\sim$(1--23)$\times 10^4$\lsol\/ in the Large and Small Magellanic Clouds using $\sim$0.1~pc-resolution 0.85~mm ALMA observations, as part of the Magellanic Outflow and Chemistry Survey (MAGOS). Using CO (3--2) emission, we identify blue- and red-shifted outflows associated with the primary continuum source toward 39 out of 40 targets. 
We derive the outflow mass, momentum, energy, and force, fit the outflow property versus luminosity scaling relations for MAGOS LMC and SMC targets, and compare them with the CO-traced outflows in the Galaxy. 
We find that the LMC CO outflows have a similar scaling relation to that in the Galaxy in mass, while the SMC CO outflows have flatter scaling relations than the LMC and the Galaxy.  
Our results are consistent with a joint dependence of CO outflows on metallicity and luminosity: substantial CO molecules in the vicinity of high-luminosity ($L\gtrsim10^{5}$~\lsol\/) MYSOs survive photodissociation at the moderately low metallicity of the LMC ($Z\sim$ 0.5~\zsol\/), while CO photodissociation becomes increasingly effective at higher luminosity (e.g., $L\gtrsim10^{4.5}$~\lsol\/) at the very low metallicity of the SMC ($Z\sim$ 0.1--0.2~\zsol\/), resulting in deficient CO outflows. 
This would suggest that the chemical behaviors of star formation processes may be regulated by the environment (i.e., metallicity and radiation field), but the physical processes in star formation (e.g., outflow launching) might remain universal across cosmic history. 
Nonetheless, the current data cannot rule out the possibility of deficient CO outflows due to intrinsically weak outflow launching. Observations of CO-dark gas are required to distinguish between CO photodissociation and intrinsically weak outflow launching in the future. 

%% Please use the acknowledgment and contribution environments. This will 
%% be anonomyized when the "anonymous" style option is used. 
\begin{acknowledgments}
We thank the referee for their constructive comments, which have greatly improved the quality and clarity of this manuscript.  
This paper makes use of the following ALMA data: ADS/JAO.ALMA\#2019.1.01770.S\@. ALMA is a partnership of ESO (representing its member states), NSF (USA) and NINS (Japan), together with NRC (Canada), NSTC and ASIAA (Taiwan), and KASI (Republic of Korea), in cooperation with the Republic of Chile. The Joint ALMA Observatory is operated by ESO, AUI/NRAO and NAOJ\@. 
This research has made use of NASA’s Astrophysics Data System. 
S.Z. and K.T. acknowledge support from the NAOJ ALMA Scientific Research Grant Code 2025-29B. 
Y.C. was partially supported by Grant-in-Aid for Scientific Research (KAKENHI number JP24K17103 and 26K00748) of the JSPS. 
This work is funded by the European Research Council (ERC) under the European Union's Horizon 2020 research and innovation program (grant agreement No. 101098309 - PEBBLES). AS acknowledges support by the grant CNS2023-145240 (Consolida). 
\end{acknowledgments}

\begin{contribution}
%%This section gives authors the space to recognize author contributions. The text inside this environment is NOT counted towards the total word quanta. At a minimum, manuscripts are expected to include this text:

SZ was responsible for performing data analysis, interpreting the results, and writing and submitting the manuscript. 
KT, YZ, and TS came up with the initial research concept. 
KT was the PI of the ALMA observations underlying this manuscript, and provided help with data analysis and interpretation. 
YZ was responsible for reducing the observed ALMA data and provided help with data interpretation and comments on the manuscript.
TS was responsible for calculating luminosities for LMC targets, and provided help with data analysis and interpretation and comments on the manuscript. 
YC provided help with data analysis and comments on the manuscript.
KF and AS provided comments on the manuscript. 

%% But authors are expected to provide more specific details, e.g. 
%%
%%SC was responsible for writing and submitting the manuscript.
%%WWM came up with the initial research concept and edited the manuscript.
%%OTS obtained the funding and edited the manuscript.
%%EBF provided the formal analysis and validation. He also edited the manuscript.
%%GEH Supervised the undergraduates, wrote the software and administers the project github and Zenodo repositories.
%%
%% Authors can use the Contributor Role Taxonomy (CRediT) at
%% for ideas on how write a good statement tailored to their needs.

\end{contribution}

\facilities{ALMA}

\software{CASA \citep{casa2022}, astropy \citep{2013A&A...558A..33A,2018AJ....156..123A,2022ApJ...935..167A}
          }


\appendix

\section{The Gallery of MAGOS outflows} \label{sec:gallery} 

\begin{sidewaysfigure*}[p!]
    \centering
    \includegraphics[
        width=\textheight,
        height=0.90\textwidth,
        keepaspectratio
    ]{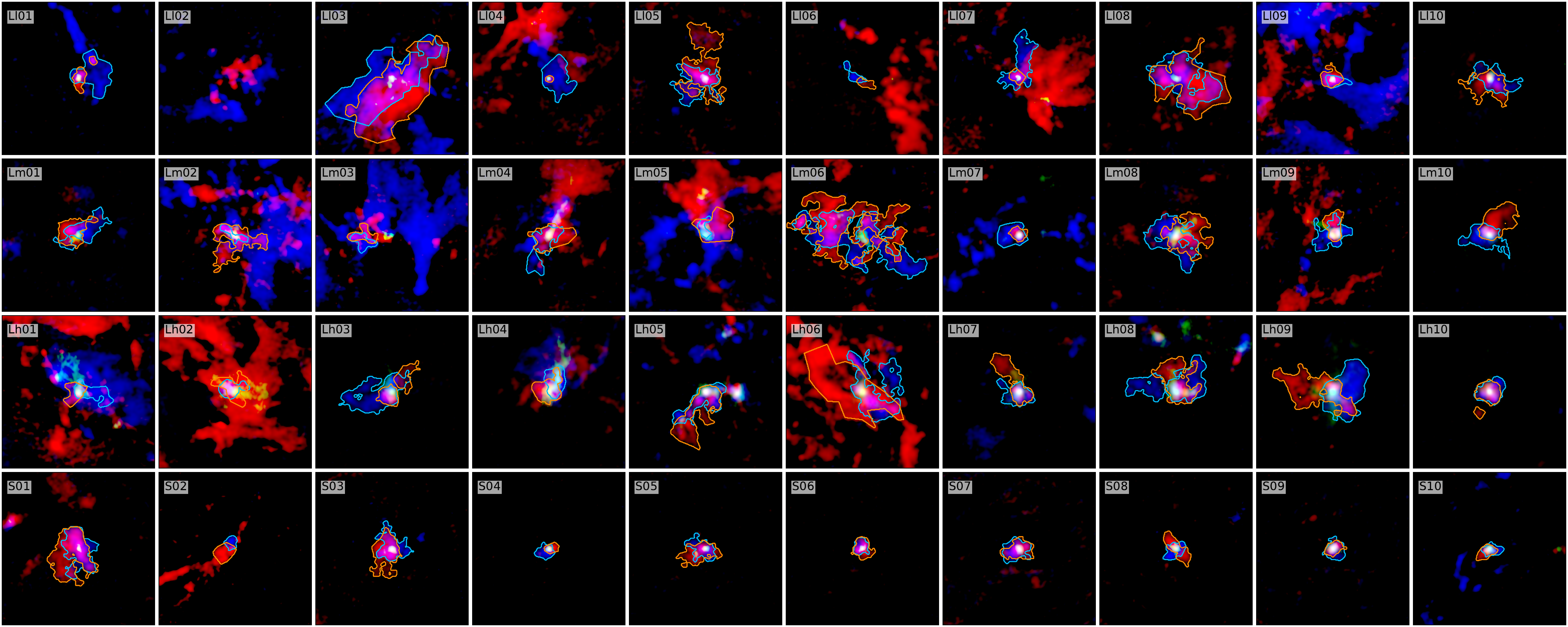}
    \caption{
    Three-color composite images of the 0.85~mm continuum emission (green) and the blue- and red-shifted integrated CO emission for the MAGOS sample. Overlaid are the outlines of determined outflow areas, with the blue-shifted lobe in blue and red-shifted in orange. 
    For Ll02, the blue and red emission is integrated over velocity ranges of [269.46, 272.97]~\kms\/ and [277.12, 278.35]~\kms\/, respectively. 
    }
    \label{fig:outflow_gallery}
\end{sidewaysfigure*}

\autoref{fig:outflow_gallery} shows the three-color composite images of the 0.85~mm continuum emission and the blue- and red-shifted integrated CO emission, overlaid with the outlines of determined outflow areas for the MAGOS sample. 
The derived outflow properties are summarized in \autoref{tab:outflow_properties}. 

\begin{splitdeluxetable*}{lcccccccccBlccccccccccBlcccc}
\tabletypesize{\scriptsize}
\tablewidth{0pt}
\tablecaption{Summary of outflow properties for the MAGOS sample. The first five targets in the LMC and SMC are shown. The full table is available in machine-readable form. 
\label{tab:outflow_properties}}

\tablehead{
\colhead{Source} &
\colhead{R.A.} &
\colhead{Decl.} &
\colhead{$L$} &
\colhead{$v_{\rm sys}$} &
\colhead{Flag\tablenotemark{a}} &
\colhead{$v_{\rm b,max}$} &
\colhead{$v_{\rm b,min}$} &
\colhead{$v_{\rm r,min}$} &
\colhead{$v_{\rm r,max}$} &
\colhead{Source} &
\colhead{$M_{\rm b}$} &
\colhead{$P_{\rm b}$} &
\colhead{$E_{\rm b}$} &
\colhead{$F_{\rm b}$} &
\colhead{$R_{\rm eff,b}$} &
\colhead{$M_{\rm r}$} &
\colhead{$P_{\rm r}$} &
\colhead{$E_{\rm r}$} &
\colhead{$F_{\rm r}$} &
\colhead{$R_{\rm eff,r}$} &
\colhead{Source} &
\colhead{$M_{\rm tot}$} &
\colhead{$P_{\rm tot}$} &
\colhead{$E_{\rm tot}$} &
\colhead{$F_{\rm tot}$} \\
\colhead{} &
\colhead{(deg)} &
\colhead{(deg)} &
\colhead{($L_\odot$)} &
\colhead{(km s$^{-1}$)} &
\colhead{} &
\colhead{(km s$^{-1}$)} &
\colhead{(km s$^{-1}$)} &
\colhead{(km s$^{-1}$)} &
\colhead{(km s$^{-1}$)} &
\colhead{} &
\colhead{($M_\odot$)} &
\colhead{($M_\odot$ km s$^{-1}$)} &
\colhead{($M_\odot$ km$^2$ s$^{-2}$)} &
\colhead{($M_\odot$ km s$^{-1}$ yr$^{-1}$)} &
\colhead{(pc)} &
\colhead{($M_\odot$)} &
\colhead{($M_\odot$ km s$^{-1}$)} &
\colhead{($M_\odot$ km$^2$ s$^{-2}$)} &
\colhead{($M_\odot$ km s$^{-1}$ yr$^{-1}$)} &
\colhead{(pc)} &
\colhead{} &
\colhead{($M_\odot$)} &
\colhead{($M_\odot$ km s$^{-1}$)} &
\colhead{($M_\odot$ km$^2$ s$^{-2}$)} &
\colhead{($M_\odot$ km s$^{-1}$ yr$^{-1}$)}
}

\startdata
Ll01 & 81.505181 & -67.503319 & $1.08\times10^{4}$ & 287.15 & S & 280.04 & 285.11 & 289.18 & 292.32 &
Ll01 & 63.81 & 155.41 & 199.83 & $1.58\times10^{-3}$ & 0.44 & 9.02 & 23.60 & 32.04 & $7.93\times10^{-4}$ & 0.22 &
Ll01 & 72.83 & 179.01 & 231.87 & $2.38\times10^{-3}$ \\
Ll02\tablenotemark{b} & 73.534502 & -66.775070 & $1.12\times10^{4}$ & \nodata & \nodata & \nodata & \nodata & \nodata & \nodata &
{} & \nodata & \nodata & \nodata & \nodata & \nodata & \nodata & \nodata & \nodata & \nodata & \nodata &
{} & \nodata & \nodata & \nodata & \nodata \\
Ll03 & 84.478456 & -69.576583 & $1.13\times10^{4}$ & 265.31 & S & 258.03 & 263.40 & 267.23 & 273.27 &
Ll03 & 431.10 & 1248.70 & 2003.41 & $1.04\times10^{-2}$ & 0.89 & 558.80 & 1862.16 & 3437.39 & $1.43\times10^{-2}$ & 0.95 &
Ll03 & 989.90 & 3110.86 & 5440.80 & $2.47\times10^{-2}$ \\
Ll04 & 75.982757 & -67.345819 & $1.37\times10^{4}$ & 268.21 & S & 261.84 & 265.63 & 270.79 & 274.54 &
Ll04 & 54.07 & 176.08 & 300.22 & $4.11\times10^{-3}$ & 0.41 & 3.02 & 10.64 & 20.55 & $1.19\times10^{-3}$ & 0.09 &
Ll04 & 57.09 & 186.72 & 320.77 & $5.31\times10^{-3}$ \\
Ll05 & 74.677027 & -66.143208 & $1.48\times10^{4}$ & 275.16 & C & 264.38 & 272.16 & 278.15 & 288.93 &
Ll05 & 68.70 & 279.30 & 611.59 & $1.18\times10^{-2}$ & 0.44 & 139.10 & 643.61 & 1669.27 & $2.68\times10^{-2}$ & 0.59 &
Ll05 & 207.80 & 922.92 & 2280.87 & $3.86\times10^{-2}$ \\
\hline
S01 & 16.378093 & -71.922569 & $1.00\times10^{4}$ & 164.88 & C & 150.11 & 162.53 & 167.22 & 176.36 &
S01 & 135.51 & 522.01 & 1165.85 & $1.21\times10^{-2}$ & 0.56 & 156.80 & 527.83 & 1028.24 & $9.97\times10^{-3}$ & 0.68 &
S01 & 292.31 & 1049.84 & 2194.09 & $2.20\times10^{-2}$ \\
S02 & 10.803773 & -72.999542 & $1.60\times10^{4}$ & 110.88 & S & 106.92 & 108.31 & 113.45 & 114.55 &
S02 & 2.80 & 8.16 & 12.05 & $1.22\times10^{-4}$ & 0.19 & 7.41 & 18.87 & 24.23 & $1.86\times10^{-4}$ & 0.30 &
S02 & 10.21 & 27.03 & 36.28 & $3.08\times10^{-4}$ \\
S03 & 11.216599 & -72.959417 & $1.90\times10^{4}$ & 142.49 & C & 132.75 & 138.86 & 146.12 & 151.38 &
S03 & 27.43 & 118.60 & 263.64 & $3.85\times10^{-3}$ & 0.45 & 62.02 & 281.96 & 662.75 & $8.81\times10^{-3}$ & 0.50 &
S03 & 89.45 & 400.57 & 926.39 & $1.27\times10^{-2}$ \\
S04 & 13.509964 & -73.355097 & $2.20\times10^{4}$ & 161.95 & C & 157.30 & 158.73 & 165.17 & 167.04 &
S04 & 5.11 & 18.12 & 32.31 & $6.21\times10^{-4}$ & 0.25 & 3.83 & 14.23 & 26.74 & $6.73\times10^{-4}$ & 0.18 &
S04 & 8.93 & 32.34 & 59.05 & $1.29\times10^{-3}$ \\
S05 & 16.455555 & -71.996792 & $2.30\times10^{4}$ & 190.30 & C & 177.62 & 186.45 & 194.15 & 208.53 &
S05 & 42.34 & 227.05 & 672.06 & $1.37\times10^{-2}$ & 0.36 & 60.62 & 312.67 & 929.04 & $1.58\times10^{-2}$ & 0.42 &
S05 & 102.96 & 539.73 & 1601.10 & $2.95\times10^{-2}$ \\
\enddata

\tablenotetext{a}{
Sources with outflow areas determined manually are flagged as S, whereas those determined using contours are flagged as C. 
}
\tablenotetext{b}{
Only the source position and luminosity are shown for Ll02.
}

\end{splitdeluxetable*}

\section{Uncertainty} \label{sec:uncertainties} 

% alpha_co factor
\subsection{CO(3--2)-to-H$_2$ conversion factor \label{sec:alpha}}

The CO(3--2)-to-H$_2$ conversion factor $\alpha_{\rm CO}$ is one of the major sources of uncertainty in the derived outflow properties. 
We first attempt to estimate $\alpha_{\rm CO}$ for each source driving outflows in the MAGOS sample. 
% how alpha factor is derived
The $\alpha_{\rm CO}$ factor is estimated by comparing the gas mass inferred from the 0.85~mm continuum flux density with the CO (3--2) luminosity measured within the same aperture. To derive the gas mass, we assume that the 0.85~mm continuum emission is dominated by optically thin thermal dust emission; 
we adopt a dust-to-gas mass ratio of 0.0027 for the LMC sources and 0.0016 for the SMC, as scaled from the Galactic value of 0.008 by their metallicities, and a dust opacity of $\kappa_{\rm 0.85~mm}=$1.97~\sqcpg\/, which is based on linear interpolation of the values for MRN grains with thin ice mantles and $n_{\rm H_2}$ = 10$^6$ \cc\/ in \cite{ossenkopf1994};  
in addition, we adopt a dust temperature of $T_{\rm dust}=$40~K for the LMC sources and $T_{\rm dust}=$45~K for the SMC, which are the typical dust temperatures of MYSOs inferred from far-infrared observations \citep{vanloon2010a, vanloon2010b}. 
The CO luminosity is derived by multiplying the CO emission integrated over the maximum blue- and red-shifted outflow velocities with the aperture area. 
A 5$\sigma$ cut-off is applied in calculating continuum flux densities and CO luminosities. 
The aperture over which the continuum flux density and CO luminosity are measured has a diameter of 0.4~pc for LMC sources and 0.3~pc for SMC sources. These diameters correspond to the typical sizes of the continuum emission associated with the primary continuum source in the LMC and the SMC, respectively. 

%source-by-source variation, LMC versus SMC
% relation to size/Tdust: coupled effect, lower Td at larger aperture
The derived $\alpha_{\rm CO}$ factors for LMC sources have a median value of 10.4 M$_\odot$ pc$^{-2}$ (K km s$^{-1}$)$^{-1}$ and an interquartile range of [7.1, 15.9] M$_\odot$ pc$^{-2}$ (K km s$^{-1}$)$^{-1}$, while those of SMC sources have a median value of 12.3 M$_\odot$ pc$^{-2}$ (K km s$^{-1}$)$^{-1}$ and an interquartile range of [5.7, 15.4] M$_\odot$ pc$^{-2}$ (K km s$^{-1}$)$^{-1}$. 
Despite a higher median value of $\alpha_{\rm CO}$ factors in the SMC, the $\alpha_{\rm CO}$ factors of the two groups largely overlap, indicating no significant dependence of $\alpha_{\rm CO}$ on metallicity in our data. 
Instead, source-by-source variations within each group can reach up to a factor of a few. These variations might be induced by uncertainties in e.g., the adopted dust temperatures, the optical depth of the dust continuum emission, and the potential contamination of free-free emission.

\subsection{Outflow scaling relations} \label{sec:stats}

% blue and red outflow property uncertainty
We first estimate the uncertainties in the derived outflow properties associated with the adopted velocity ranges. Following \cite{maud15b}, we perturb the lower and upper channel boundaries of each blue- and red-shifted outflow lobe by $\pm1$ channel ($\sim$0.4~\kms\/) and recalculate the outflow properties. 
%
% total outflow property
We then derive the uncertainty of each total outflow property directly in logarithmic space from all combinations of independently perturbed blue- and red-shifted outflow velocity ranges relative to the nominal case.  
The median rms values are 0.07, 0.05, 0.04, and 0.02~dex for mass, momentum, energy, and force, respectively, with maximum values of 0.29, 0.24, 0.21, and 0.22~dex. 
% MAGOS luminosity
For LMC targets, the uncertainty of the LMC distance is $\sim$2\% \citep{pietrzynski2013} and the uncertainty of flux measurements and integration is estimated to be 10\%. Thus the luminosity uncertainty is taken to be 14\%. 
For SMC targets, the luminosity uncertainty is taken to be 11\%, the median fractional uncertainty reported by \cite{sewilo2013}. 

These uncertainties are considered minimum uncertainty estimates for the derived outflow properties and luminosities of the MAGOS sample and are adopted in the linear regression analysis. 
% MW uncertainties
For the Galactic sample, fractional uncertainties of 36\%, 26\%, 23\%, 50\%, and 30\% are adopted for mass, momentum, energy, force, and luminosity, respectively \citep{maud15a, maud15b}. 

% other sources of uncertainties 
We note that there are other sources of potentially significant uncertainty. 
% inclination angle to outflow properties
First, the inclination angles of outflows can introduce uncertainties into velocity-dependent outflow properties such as momentum, energy, and force. 
For an inclination angle $\theta$ with respect to the line of sight, the intrinsic momentum, energy, and force can be obtained by multiplying their observed values by factors of $1/\cos{\theta}$, $1/\cos^2{\theta}$, and $\sin{\theta}/\cos^2{\theta}$, respectively. 
For a representative inclination angle of $\theta=57.3^{\circ}$ \citep[e.g.,][]{bontemps1996}, correction factors of 1.85, 3.43, 2.89 should be applied to observed momentum, energy, and force, respectively.   
% inclination angle to luminosity
Moreover, \cite{zhang2014, zhang2018} suggest that the flashlight effect due to different viewing angles is an important source of uncertainty in the estimated luminosity and can induce a difference of a factor of a few between estimated and intrinsic luminosities, depending on evolutionary stages. 
However, with a spatial resolution of $\sim$0.1~pc, our MAGOS data are insufficient to constrain the inclination angle source by source, which is also the case for the Galactic reference sample in \cite{maud15b}. 
Applying a uniform correction factor would change only the normalization but not the slopes of the scaling relations.
Therefore, we do not apply inclination corrections to the derived outflow properties and luminosities of the MAGOS sample and compare them with the uncorrected values of Galactic outflows. 
% multiplicity
Second, for some targets with resolved multiple compact continuum sources (e.g., Lm04, Lh05), the estimated luminosity may include contributions from multiple sources and can thus be overestimated for the primary source. 
% possibility of unresolved clusters of outflows
% e.g., Lh09
In particular, even a single continuum source in the 0.1~pc-resolution data may be resolved into a cluster of compact sources with multiple outflows in higher-resolution observations \citep[e.g., Lh09,][]{cheng2026}. 
In this case, the MAGOS data captures a cluster of outflows rather than outflows driven by a single MYSO. 

To assess the robustness of our results to the adopted uncertainty assumptions, we perform linmix linear regression on the outflow-luminosity scaling relations for the LMC, SMC, and the Galaxy separately by considering (1) no uncertainty and (2) uniform uncertainties of 0.2~dex, 0.3~dex, and 0.5~dex for all properties ($M$, $P$, $E$, $F$ and $L$) of MAGOS targets. We find that the main conclusions remain unchanged when a uniform uncertainty of $\leq$0.2~dex is adopted, although the SMC relations become less constrained with increasing uncertainties. With 0.5~dex uncertainties, the SMC relations and the LMC $M$-$L$ relation are unconstrained while the LMC still has steeper slopes than the Galaxy in $P$, $E$, and $F$. These results suggest that assigning a uniformly large uncertainty to all properties does not unveil additional physical information, but instead yields weaker constraints. 

% subsample tests
% questions:  whether a true slope difference exists but is undetectable with N=10?
In addition, we examine whether the deviations in the LMC and SMC outflow scaling relations could arise purely from their small sample sizes and limited luminosity ranges. 
First, we perform subsampling of the Galactic sources with luminosities of $L\geq10^4$~\lsol\/ (see \autoref{sec:obs}) with the sample size matched to the LMC (N=29) and the SMC (N=10) and refit each subsample using the linmix algorithm. 
This process is repeated 1000 times for each group and each outflow property. 
The distributions of posterior median slopes of mock Galactic samples with LMC conditions are Gaussian-like and are fitted with Gaussian models, while the distributions with SMC conditions have long tails. 
As summarized in \autoref{tab:subsampling}, the slopes are steeper and have larger scatter when smaller sample sizes and narrower luminosity ranges are adopted. 
For the LMC, while the posterior median slope is consistent with the posterior median slopes of mock Galactic samples with LMC conditions within 1$\sigma$ in $M$, it becomes progressively steeper from $M$ to $F$. 
For the SMC, the posterior median slope is below the central 68\% interval of the posterior median slopes of mock Galactic samples with SMC conditions in all properties, although their central 68\% intervals increasingly overlap with each other from $M$ to $F$. 
The intercepts, defined as the normalization of the scaling relations at $L=10^4$~\lsol\/, are systematically higher for the MAGOS sample than for the Galactic sample. 
The Galactic mock samples yield slightly lower intercepts than the Galactic sample, with very similar median values for the LMC- and SMC-like mock samples; this indicates that the imposed luminosity selection has a modest effect on the normalization, while reducing the sample size mainly increases its scatter. Compared with these Galactic mock distributions, the observed LMC and SMC posterior-median intercepts are higher for all four outflow properties.  
The excess is most pronounced for the LMC mass and momentum relations, while the broad SMC posterior intervals lead to substantial overlap with the corresponding mock distributions.

\begin{deluxetable*}{llccccccccc}
\tabletypesize{\scriptsize}
\tablecaption{Observed and mock regression parameters of the outflow-luminosity scaling relations.
\label{tab:subsampling}}
\tablewidth{0pt}
\tablehead{
\colhead{Property} &
\colhead{Parameter} &
\colhead{Galaxy} &
\colhead{Galaxy 68\%} &
\colhead{LMC} &
\colhead{LMC 68\%} &
\colhead{Mock LMC $\mu\pm\sigma$} &
\colhead{SMC} &
\colhead{SMC 68\%} &
\colhead{Mock SMC} &
\colhead{Mock SMC 68\%}
}
\startdata
Mass
 & $\alpha$ & 1.50 & (1.45, 1.54)
 & 1.88 & (1.69, 2.05) & $1.45\pm0.06$
 & 1.92 & (1.52, 2.29) & 1.44 & (1.20, 1.66) \\
 & $\beta$ & 0.61 & (0.50, 0.71)
 & 0.69 & (0.47, 0.91) & $0.64\pm0.11$
 & $-0.37$ & ($-1.10$, 0.34) & 0.67 & (0.30, 1.09) \\
\hline
Momentum
 & $\alpha$ & 2.07 & (2.02, 2.12)
 & 2.30 & (2.08, 2.53) & $1.99\pm0.07$
 & 2.46 & (1.99, 2.91) & 1.97 & (1.76, 2.18) \\
 & $\beta$ & 0.75 & (0.64, 0.85)
 & 1.10 & (0.83, 1.36) & $0.87\pm0.12$
 & $-0.26$ & ($-1.11$, 0.61) & 0.88 & (0.56, 1.33) \\
\hline
Energy
 & $\alpha$ & 2.74 & (2.67, 2.80)
 & 2.77 & (2.49, 3.03) & $2.63\pm0.07$
 & 3.01 & (2.46, 3.51) & 2.62 & (2.35, 2.87) \\
 & $\beta$ & 0.89 & (0.75, 1.02)
 & 1.54 & (1.20, 1.86) & $1.06\pm0.13$
 & $-0.11$ & ($-1.04$, 0.87) & 1.06 & (0.66, 1.56) \\
\hline
Force
 & $\alpha$ & $-2.65$ & ($-2.71$, $-2.59$)
 & $-2.51$ & ($-2.78$, $-2.24$) & $-2.79\pm0.08$
 & $-2.21$ & ($-2.77$, $-1.66$) & $-2.78$ & ($-3.03$, $-2.56$) \\
 & $\beta$ & 0.69 & (0.57, 0.80)
 & 1.64 & (1.30, 1.96) & $0.98\pm0.13$
 & 0.28 & ($-0.74$, 1.26) & 0.96 & (0.55, 1.34) \\
\enddata
\tablecomments{
$\alpha$ and $\beta$ are the posterior median intercepts and slopes, respectively, of the LINMIX linear regression, with their central 68\% posterior credible intervals. $\alpha$ represents the normalization at $L=10^4\,L_\odot$. 
For the Galactic mock samples with LMC conditions, $\mu$ and $\sigma$ are the mean and standard deviation of the best-fitting Gaussian model for the distribution of posterior median regression parameters. 
For the Galactic mock samples with SMC conditions, the median of the posterior median regression parameters and the central 68\% interval of the mock distribution are shown instead. 
}
\end{deluxetable*}

%% For this sample we use BibTeX plus aasjournalv7.bst to generate the

\bibliography{main}{}
\bibliographystyle{aasjournalv7}


\end{CJK*}
\end{document}